# Effects of Trailing Edge Thickness on NACA 4412 Airfoil Performance at Low Reynolds Numbers: A CFD Analysis

Sayed Tanvir Ahmed[a)] and Mahadi Hasan Shanto

*Department of Mechanical Engineering, Shahjalal University of Science and Technology, Sylhet-3114, Bangladesh*

[a)] Corresponding author: sytanvir.mech@gmail.com

**Abstract.** Due to the augmentation of the significance of wind energy, giving a high priority to the airfoil's efficiency enhancement is obligatory. To improve the performance of airfoils, many impressive techniques are already invented. In this article, the trailing edge of the NACA 4412 airfoil is modified by changing the thickness. CFD is used in this study, which aids in the identification of several important details. For our investigation, we choose the reliable Spalart Almaras model and the Reynolds number is 300k. Overall, the results demonstrate that using 0.8% thickness at the trailing edge of the NACA 4412 airfoil is viable to obtain the best output. The predominant reason is that not only the better coefficient of lift but also the preferable lift-to-drag ($C_L/C_D$) ratio is found in this configuration. However, using 0.2% thickness at the trailing edge reduces performance as a whole. So, it is recommended to utilize 0.8% thickness on the trailing edge of the NACA 4412 airfoil.

## 1. INTRODUCTION

In this contemporary era, the demand for renewable energy is proliferating and without focusing on renewable energy the world will hardly survive in the near future. There are some specific reasons why the world is focusing on renewable energy. One of the most predominant causes is that the quantity of fossil fuels (coal, natural gas, and oil) is diminishing at a rapid pace, as the energy demand is increasing swiftly. Wind energy is considered one of the most potential solutions. Many Governments around the globe are investing highly in wind energy. Despite solar energy being a potential option to solve the future energy crisis, the extraction of solar energy is not feasible in all seasons. In addition, geothermal power is an excellent source of renewable energy, however, there are few reliable sources available. This type of energy completely depends on the temperature of the sources, but the maximum sources only provide 150°C or less than 150°C which is not ideal (Li et al., 2020). Ocean is a good source of power but the problem is that the location is very restricted. Only some countries around the globe will be able to extract this type of energy. And also, due to the high cost, many nations are not motivated to concentrate on ocean energy. On the contrary, wind energy is suitable for most countries in the world, as it has good feasibility and cost-effectiveness. Transforming wind energy into electrical energy is the main moto of wind turbines. The wind turbine's blade is a major part since it plays a crucial role to determine the efficiency of a wind turbine. The cross-sectional of the blade is termed the airfoil. In order to magnify the overall efficiency of the wind turbine, focusing on the performance of the airfoil is obligatory. So, if we can determine the best-performed airfoil, the power output maximization of the wind turbine will be expected. Studies on airfoils are ongoing in an effort to identify the most efficacious airfoil. Researchers have already completed several superb works on the airfoil, which provide fruitful performance. Computational Fluid Dynamics(CFD) has been utilized to realize numerous significant characteristics of flow fields. In other words, CFD has been utilized to comprehend the airfoil's performance. There are some important factors that can help to improve the performance of the airfoil. Using a flap, vortex generator, Co flow jet control, slotting, geometrical modification, and cavity aid to magnify the performance of the airfoil. Trailing edge thickness can be another impressive way that can boost the performance of the airfoil substantially.

The utilization of wind turbines has been augmented throughout the last two decades. For instance, in the USA, a total of 1800 wind turbine power plants having a capacity of 2454 MW were built in 2006, but in 2009, a dramatic change was noticed in the US market because they installed 5760 units having a capability of 9922 MW (Michalak & Zimny, 2011). Even, in recent years, this type of augmentation of wind turbines is noticeable in the US market. In this day and age, a considerable increment rate of wind turbine installation is noticed the Indian subcontinent. In Bangladesh, the power generation of wind energy rose from 80 MW in 2021 to 120 MW in 2022, which indicates a 50% increment (Khare et al., 2022). When it comes to talking about the performance of wind energy, unfortunate information that the overall performance of the wind turbine is not satisfactory. The optimum efficiency of the horizontal axis wind turbine is 59.3% (Schubel & Crossley, 2012). Researchers are applying numerous techniques so that they can enlarge the performance of wind turbines. The wind turbine's aerodynamic performance mostly depends on the airfoil's performance. There are numerous ways that help to raise the airfoils' performance. A study on the NACA 4412 airfoil using flap in extreme ground effect suggests that the small flap deflection on the NACA 4412 airfoil helps to boost the $C_L$-$C_D$ ratio (Ockfen & Matveev, 2009). Static vortex generator helps to increase the maximum coefficient of lift and stall angle at a lower Reynolds number as like as the higher Reynolds number (Seshagiri et al., 2009).

If slots are used on the airfoil at the perfect position, an airfoil can provide a better coefficient of lift and drag ratio. The NACA 4412 airfoil's local pressure is enhanced when the leading edge is slotted, which raises the lift coefficient and improves overall aerodynamic performance (Beyhaghi & Amano, 2017). Active flow control by Co Flow Jet (CFJ) is another impressive way, which improves the overall performance of airfoils. An oscillating SC1095 airfoil's simulation with Co flow jet flow demonstrates that the aerodynamic performance improves in every flow (Lefebvre & Zha, 2013). Utilizing a cavity in the airfoil could be effective to raise the performance. A study on NACA 0018 airfoil using a cavity suggests that the coefficient of lift and drag ratio improves at some certain degree angles of attack (AOA). (Lam & Leung, 2018).

In this work, the NACA 4412 airfoil is modified by changing the trailing edge thickness. To determine the influence of the trailing edge thickness on the NACA 4412 airfoil, computational fluid dynamics (CFD) was performed on the airfoil. In our work, Ansys Fluent, which is a CFD simulation software, is used to obtain the results. A 300k Reynolds number was used for the investigation. This study will help us to choose the best trailing edge thickness in the NACA 4412 airfoil, which will provide the optimum performance.

## 2. METHODOLOGY

A body experiences both lift force and drag force as it remains in the air. The coefficients of lift and drag represent the lift force and drag force respectively and those two are regarded as the foremost parameters to comprehend the airfoil's aerodynamic performance. The coefficient of lift ($C_L$), which is demonstrated in equation 1, is a dimensionless quantity made up of the variable's lift force $F_L$, density $\rho$, wind speed $V$, and chord length $c$. Similarly, the drag coefficient ($C_D$), which is determined using Eqn. 2, is a dimensionless quantity that represents the drag force $F_D$. The distance between the trailing edge and the leading edge of an airfoil is termed the chord length. Another crucial element in Equation 3 is the pressure coefficient $C_p$, where the pressure difference is denoted by $\Delta P$. The relative pressure throughout a flow field is indicated by the pressure coefficient.

$$C_L = \frac{F_L}{2\rho V^2 c} \qquad (1)$$

$$C_D = \frac{F_D}{2\rho V^2 c} \tag{2}$$

$$C_p = \frac{\Delta P}{2\rho V^2} \tag{3}$$

## 2.1 Numerical Methods

Indisputably, Computational Fluid Dynamics (CFD) is omnipresent where the fluid flow is the major focal point. In this triumph, to understand a fundamental fluid dynamics problem, the CFD method is utilized to simulate the flow around airfoils. The Navier-Stokes equation which describes the flow of fluid is considered the base of the CFD simulation. The conservation law governing the physical characteristics of fluids forms the basis of this equation. Specifically, mass, momentum, and energy conservation are the base of this equation. The mass and momentum conservation equations are solved for all flows using the reliable platform Ansys Fluent. Eq. 4 and Eq. 5 represent the mass conservation and momentum conservation equations respectively (Tu et al., 2018).

$$\frac{\partial \rho}{\partial t} + \nabla \cdot (\rho \vec{V}) = S_M \tag{4}$$

$$\frac{\partial}{\partial t}(\rho \vec{V}) + \nabla \cdot (\rho \vec{V} \vec{V}) = -\nabla p + \nabla \cdot (\bar{\bar{r}}) + \rho \vec{g} + \vec{F} \tag{5}$$

$$\bar{\bar{r}} = \mu[(\nabla \vec{V} + \nabla \vec{V}^T) - \frac{2}{3}\nabla \vec{V} I] \tag{6}$$

According to the mass conservation equation, $\vec{V}$ is as a velocity vector. In addition, the velocity vector is the function of position (x, y, z) and time. After that, the position has velocity components (u, v, and w) and $S_M$ is considered as the source term. In the conservation of momentum equation, $\vec{V}$ is the velocity vector as stated before, static pressure is denoted by p, $\rho \vec{g}$ is the gravitational body and $\vec{F}$ is the external body force. And finally, $\bar{\bar{r}}$ is termed a stress tensor. The equation of $\bar{\bar{r}}$ is written in Eqn. 6 where viscosity is $\mu$ and the unit tensor is indicated by I.

The Spalart-Almaras model is a single-equation model, which is specially designed for aerospace-related applications. A modeled transport Equation for kinematic eddy viscosity is solved by using this model. In this model, $\tilde{v}$ is considered the variable, which is not affected by the strong viscous effects. The equation that is given below has the capability to treat turbulent flow when the body is in a laminar region.

$$\frac{\partial}{\partial t}(\tilde{v}) + \frac{\partial}{\partial t}(\rho\tilde{v}u_i) = Gv + \frac{1}{\sigma\tilde{v}}\left[\frac{\partial}{\partial x_j}\left\{(\mu + \rho\tilde{v})\frac{\partial\tilde{v}}{\partial x_j}\right\} + Cb2\rho\left(\frac{\partial\tilde{v}}{\partial x_j}\right)^2\right] - Y_v + S_{\tilde{v}}$$

In this equation, $Gv$ indicates the turbulent viscosity production and $Y_v$ represents the destruction term. The other factors utilized in this method are constant.

The Spalart-Allamaras model expects that mesh is sufficiently refined closer to the wall surfaces having non-dimensional wall distance $y^+$ is defined in terms of friction velocity $u_*$ as denoted in Eqn. 7 and Eqn. 8; whereas the $\tau_\omega$ the term is the wall shear stress. The flow is laminar the more it gets closer to the wall hence $\tilde{v}$ is set to zero.

$$y^+ = \frac{u_*}{v} \tag{7}$$

$$u_* = \sqrt{\frac{\tau_\omega}{\rho}} \tag{8}$$

## 2.2 Flow Domain and Grid Generation

The initial preprocessing of the simulation setup is crucial as a one-minute error can result in drastically wrong decisions. Boundary conditions are the most significant initial preprocessors as they direct the flow variables of the designed physical model. Therefore, it is a matter of great attention that assigned boundary conditions namely, inlet, outlet, wall, and interface be consistent. In Table 1, some important boundary conditions are mentioned.

**Table 1:** Initial Boundary Conditions of the CFD Analysis

| Simulation Property | Parameters | Solver Type | Time | Viscous Model | Number of Iteration | Momentum | Pressure Velocity Coupling |
|---|---|---|---|---|---|---|---|
| | Value | Pressure based | Steady | Spalart-Allmaras | Close to 1000 | Second-order upwind | Simple |

| Fluid Property | Parameters | Fluid | Density (kg/m³) | Viscosity (kg/m-s) | Angle of Attack | Reynold Number | Pressure |
|---|---|---|---|---|---|---|---|
| | Value | Air | 1.225 | 1.8E$^{-5}$ | 0° to 20° | 300000 | 1 atm |

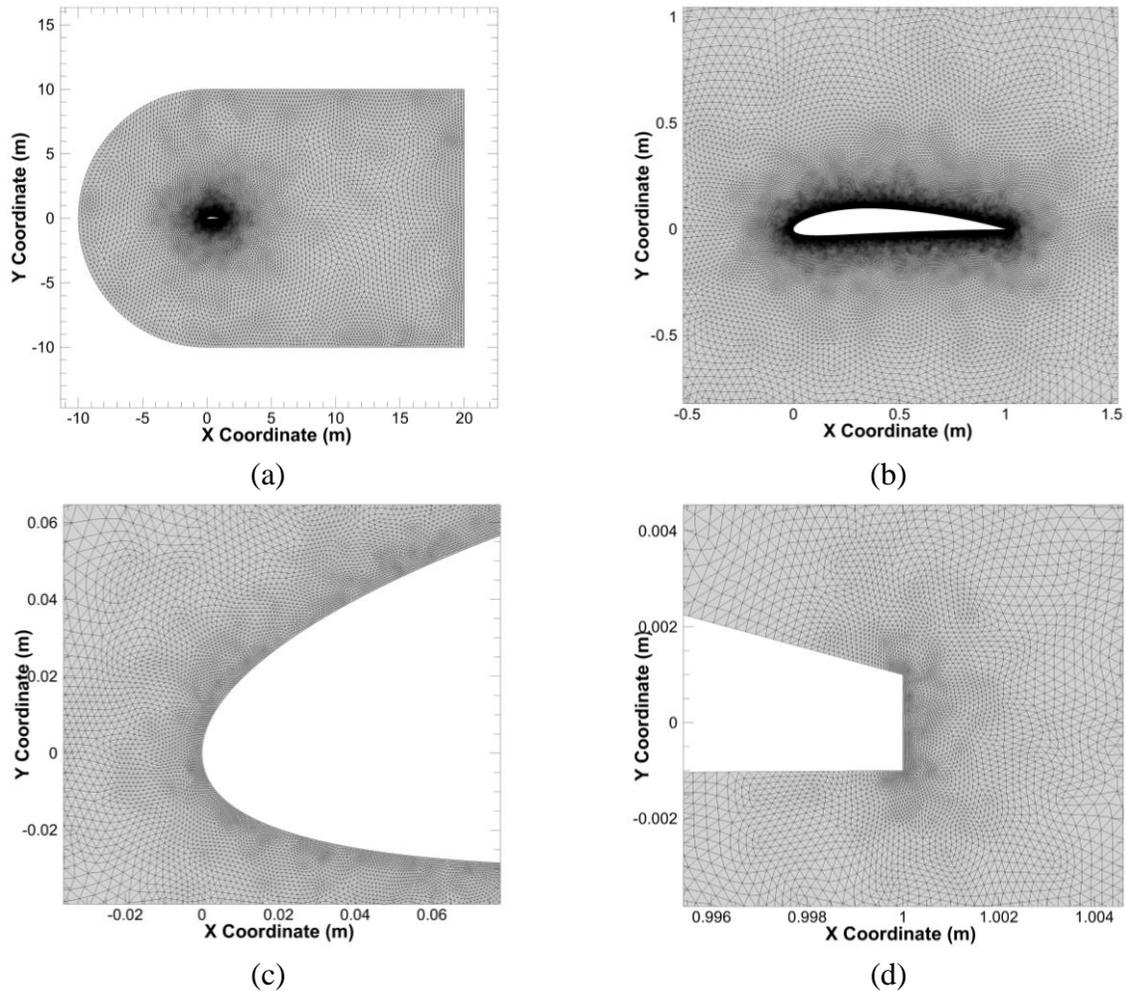

(a)  (b)

(c)  (d)

**Fig. 1.** C-Type flow domain

The flow domain is C-type, where a semicircle having a 10m radius is utilized. The width of the rectangular domain part is 20m. At the center, the airfoil is located as shown in Fig.1. The velocity inlet as inlet initial boundary conditions have been specified. Furthermore, the no-slip wall condition is set to the airfoil. By using the Reynolds number we find the inlet velocity, which is 4.26 ms$^{-1}$. At 1 atm pressure with 15°C temperature the dynamic viscosity remains $\mu =$

$1.8 \times 10^{-5}$ kg/ms. The mesh's level of smoothness has a significant impact on how well the simulation performs. The wall y+ is kept at a certain level so that the flow can be predicted with good accuracy. Refinement is used finally to obtain the most accurate results. The number of nodes is around 85 thousand and the elements number is around 0.17 million for all the models.

## 2.3 Validation of CFD Model

In order to measure the accuracy of the turbulence model and the overall generated result of the paper, we have compared lift coefficient curves with experimental and numerical data developed by Eleni (Douvi C. Eleni, 2012). In the research work, the Reynolds number was 30,00,000 at 300K temperature having 1.225 kg/m$^3$ density and 1.8×10$^{-5}$ Ns-m$^{-2}$. On NACA 0012 airfoil from

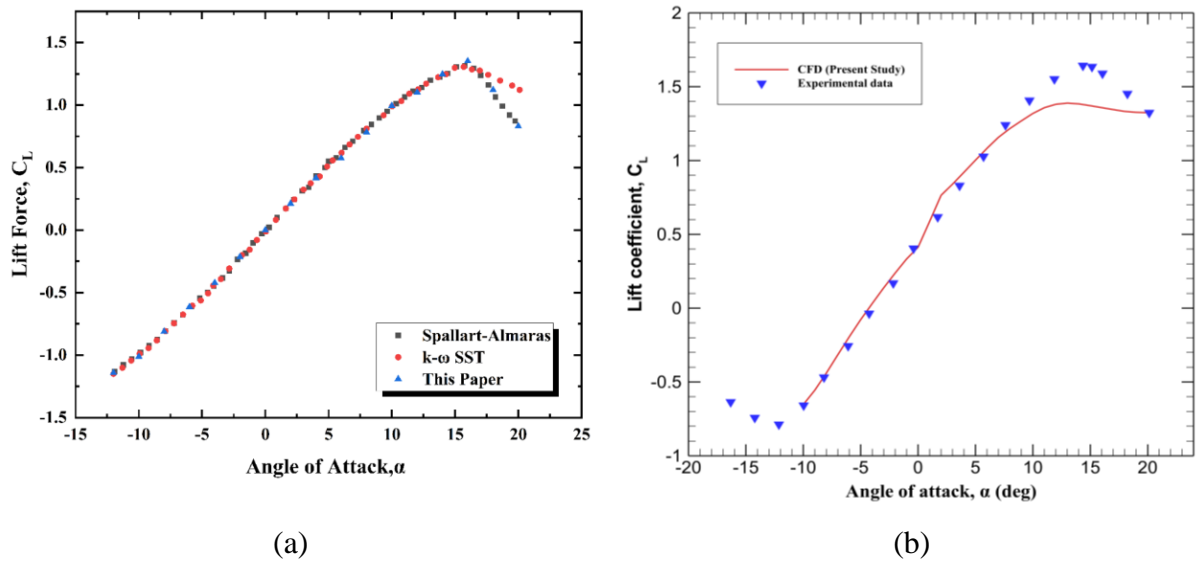

(a)          (b)

**Fig. 2.** Comparison of numerically predicted lift coefficient with experimental data of (a) Eleni (Douvi C. Eleni, 2012) and (b) (Abbott, 1945).

−12° to 20° angle of attack was experimented and simulated. Abbott's (Abbott, 1945) experiments and our results provide further support for the study. In this work, we used numerical techniques to carefully assess the experimental data. Our results' degree of agreement is shown in Figure 2. The experimental and numerical results agreed well, with negligible error. This enhances our study's dependability and offers a strong foundation for more investigation.

## 2.4 Airfoil Conceptualization

NACA 4412 airfoil is one of the most used airfoils in the area of wind engineering. A 1000 mm NACA 4412 airfoil is taken in this study using 100 coordinates. The letters NACAYYXX stand for the family of NACA airfoils, where the percentage of the maximum camber is indicated by the first digit, the position of the maximum camber from the leading edge in tenths of the chord is denoted by the second digit, and the maximum thickness to chord ratio is represented by the final two digits. The following equation was used to create the airfoil:

$$y_t = 5t\left[0.2969\sqrt{\frac{x}{c}} - 0.1260\left(\frac{x}{c}\right) - 0.3516\left(\frac{x}{c}\right)^2 + 0.2843\left(\frac{x}{c}\right)^3 - 0.1015\left(\frac{x}{c}\right)^4\right]$$

Where the half-thickness is represented by $y_t$ at a particular value of x (centerline to the surface), c indicates the value of the chord length, the point x is situated along the chord from 0 to c, and the maximum thickness t stays in the equation as a percentage of the chord. For our study, we have mainly modified the trailing edge of the NACA 4412 airfoil. We have developed 11 configurations. The trailing edge thickness in the first scenario is 0%, where there is no gap on the trailing edge. The trailing edge thickness is 0.1% in the second and 0.2% in the third case. By doing this, we have moved forward to 1% within a 0.1% interval.

## 3. RESULT AND DISCUSSION

In order to determine the best- and worst-performing airfoils, the results of NACA 4412 airfoil considering trailing edge thickness under various angles of attack (0° to 20°) are reported in this section. The airfoil's performance with the angle of attack is first examined in connection to one of the most crucial parameters, the coefficient of lift. Another major parameter, the drag coefficient is introduced next. Additionally, the lift-to-drag ratio is examined to determine which thickness of NACA 4412 airfoil demonstrates the best and worst performance. Following that, in an effort to pinpoint the underlying cause of such performance, static pressure contours are analyzed. Finally, to determine the velocity distribution around the airfoils, the contours of velocity are displayed for the best, the baseline (where the trailing edge thickness is 0%), and the worst.

## 3.1 Investigation of Lift Coefficient ($C_L$) with the Angle of Attack (AOA)

It is well known that the lift coefficient is one of the most predominant variables in the study of Airfoil. The angle of attack has a strong impact on the lift and drag coefficient acting on the airfoil. In our study, in general, it is found that the coefficient of lift rises to a particular angle of attack, however, passing the particular angle, this quantity starts to decline until it reaches 20° AOA. This particular angle is called the stall angle, where the stall angle refers to the angle of attack at which the coefficient of lift is the greatest. In our study, the stall angle ranges are generally from 12° to 14° for almost all cases.

When there is no trailing edge thickness considered, we find that the value of the $C_L$ is 0.39581 at 0° AOA. This value enhances until it reaches 14°. The highest value, also known as the stall angle for this circumstance, is 1.3892 at a 14° AOA. The coefficient of lift starts to diminish after crossing this stage, reaching a value of 1.0989 at a 20° angle of attack. The trend is nearly the same when trailing edge thickness is taken into account. At the configuration of 0.8% trailing edge thickness, the coefficient of lift at 0°AOA is 0.42061. For this condition, the highest value of $C_L$ is 1.4053, which is discovered at the 14° angle of attack. Intriguingly, this is the highest stall angle value for all cases in our study.

The optimal $C_L$ vs. AOA curve is discovered at 0.8% trailing edge thickness after taking into account every condition in our investigation. On the other hand, at the 0.2% trailing edge thickness configuration, this scenario is the worst. In other words, the 0.2% trailing edge thickness demonstrates the worst $C_L$ vs AOA curve among all configurations. In this case, the value of $C_L$ is 1.2774 at a stall angle of 12°, which indicates a weak stall angle value. In brief, the $C_L$ vs AOA for the total eleven configurations in our study proclaims that the 0.8% trailing edge thick NACA 4412 airfoil provides the best $C_L$ vs AOA curve, on the contrary, the 0.2% trailing edge thick configurations denotes the worst scenario. The following diagram provides the $C_L$ vs AOA curve for 0.8% trailing edge thick and 0.2% trailing edge thick airfoil compared with the 0% thick airfoil.

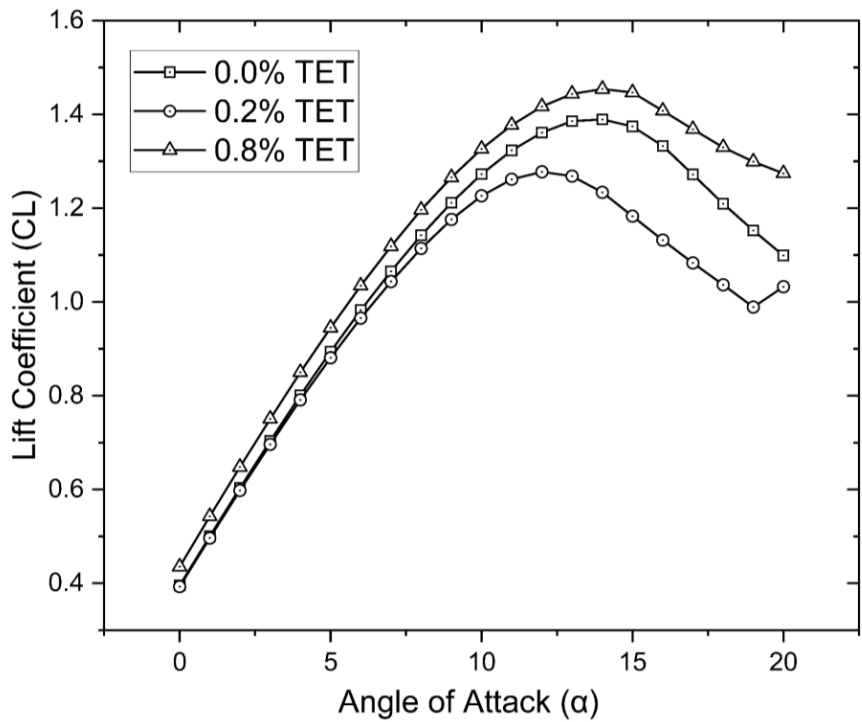

**Fig. 3.** Coefficient of lift with the angle of attack for 0%, 0.2%, 0.8% trailing edge thickness.

### 3.2 Investigation of Drag Coefficient $C_D$ with the Angle of Attack (AOA)

Despite the airfoil's acceptable lift performance giving a huge idea about the overall performance of the airfoil, the complete picture can only be seen by examining the drag performance. How well an aerodynamic body goes forward by minimizing air resistance is determined by the drag coefficient.

In every case of our study, the drag coefficient is still within acceptable limits at the 0° to 14° or 15° AOA. At this stage, the increment of the $C_D$ is low. However, after passing through this stage, this value starts to increase rapidly. At the higher angle of attack, the value of the drag coefficient is dramatically higher. When the trailing edge thickness of the NACA 4412 airfoil is 0%, the value of the $C_D$ is 0.0145 at 0° AOA. Although this value increases gradually till the 15° angle of attack, a dramatic augmentation is noticed after crossing this stage. The value rises from 0.059 at 15° AOA to 0.073 at 16° AOA, which indicates a huge increment. This increment is continued until it reaches 20° AOA. At 0.8% trailing edge thick configuration, the same scenario is found. In this case, the coefficient of drag is 0.0145 at 0° AOA and the increment is continued till 15° AOA, where the value is 0.06. It indicates a steady increment. However, after crossing 15°, the increment is significant. The value of $C_D$ rises from 0.06 at 15° AOA to 0.174 at 20° AOA, which indicates a 289% increment. The $C_D$ vs. AOA curve of 0.8% and 0.2% trailing edge thickness in comparison to the 0% trailing edge thickness configuration is given in the diagram below.

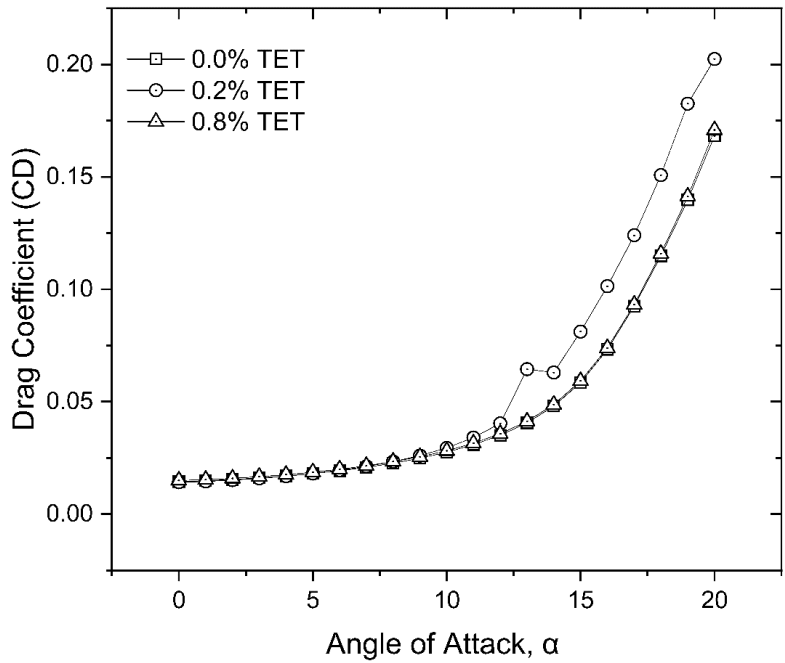

**Fig. 4.** Coefficient of drag with angle of attack for 0%, 0.2%, 0.8% trailing edge thickness.

## 3.3 Investigation of Coefficient of Lift and Coefficient of Drag Ratio with the Angle of Attack (AOA)

Irrefutably, the ratio of the coefficient of lift and coefficient of drag is the most salient factor to find the best performance. The optimal $C_L/C_D$ vs. AOA curve is obtained at 0.8% trailing edge thickness of the NACA 4412 airfoil. At the 0% trailing edge thickness configuration of the airfoil, the $C_L/C_D$ value is 27.39549 at the 0° angle of attack. This figure continuously rises until the angle of attack reaches 7°. In this position, the value of $C_L/C_D$ is 51.18167 and this is the highest value for this condition. However, after passing the 7° angle of attack, this value starts to diminish till 20°. At 20° AOA, the value of $C_L/C_D$ is 6.529412, which is the lowest in this case. The major reason behind the lower $C_L/C_D$ value at the higher angle of attack is the higher value of the drag coefficient. When it comes to the term of 0.8% trailing edge thickness situation, the trend is the almost same as the 0% thickness condition. But, overall performance in this case is better compared to other cases.

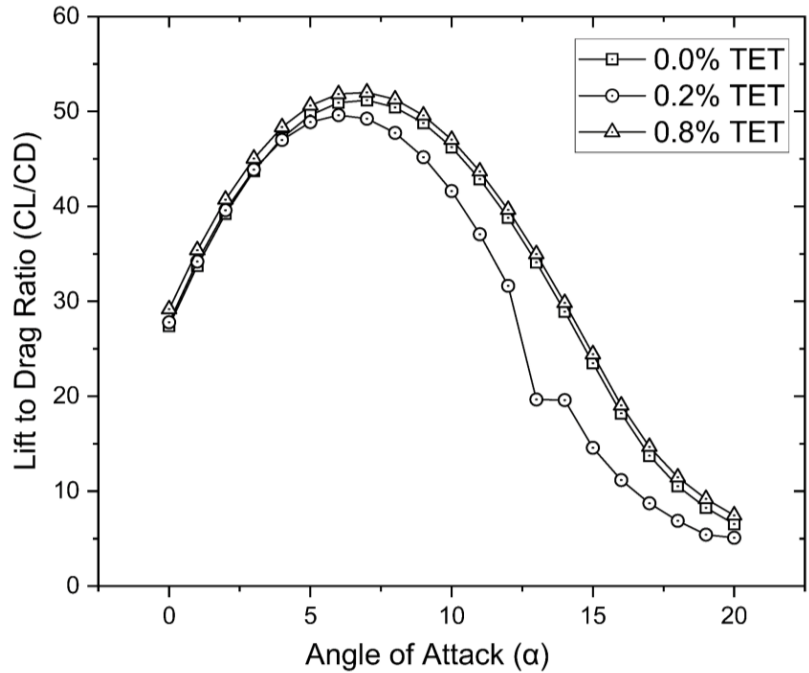

**Fig. 5.** Lift-to-drag ratio with angle of attack for 0%, 0.2%, 0.8% trailing edge thickness.

In this case, the value of $C_L/C_D$ is 28.16459 at 0° angle of attack. This number is rising steadily, reaching its greatest point of 50.98652 at a 7° AOA. Following that, this number begins to decline, with the worst-case situation being discovered at a 20° angle of attack, which is 6.455407.

The $C_L/C_D$ vs AOA curve for this condition demonstrates the better curve among all conditions. On the 0.2% trailing edge thickness, the worst scenario is evident. At almost every angle of attack, the value of $C_L/C_D$ is low compared to the other conditions. The value of the lift coefficient and drag coefficient ratio is 27.80048 at the 0° angle of attack, which is the lowest among all initial conditions. The highest ratio is 49.60791 at the 6° angle of attack, which is not a satisfactory value. The following graph demonstrates the $C_L/C_D$ vs AOA scenario for the best and worst configurations compared with the 0% thickness condition.

## 3.4 Investigation of Static Pressure Contour, Velocity Pressure Contour and Pressure Coefficient

Any aerodynamic body's wings provide lift by developing more pressure at the lower than the upper parts of the body, providing a net upward force. Now, if the pressure below the airfoil rises due to a greater angle of attack, the net force rises, resulting in greater lift, but also greater total drag. Drag is developed when the pressure of the leading edge is greater than the pressure of the trailing edge of the airfoil. From the previous section, we have found the best-and worst performed airfoil, which is 0.8% and 0.2% trailing edge thick airfoil respectively. The following diagram demonstrates the static pressure contour of the NACA 4412 airfoil with 0% trailing edge thickness, 0.2% trailing edge thickness, and 0.8% trailing edge thickness. Intriguingly, it can be seen from the diagram that the airfoil's pressure difference between the upper and lower edge increased with the increment of the angle of attack, however, after passing a certain level, this difference starts to drop. At 0% thickness of NACA 4412 airfoil, it is found that the pressure difference is relatively low at 0° AOA, however, this value starts to enhance till the stall angle. At the stall angle, the pressure difference is the maximum, for which, the higher coefficient of lift is found. After that, the difference starts to reduce until it reaches the 20° AOA. While looking at the drag coefficient for this condition, it is found that the pressure difference between the leading edge and trailing of the airfoil enhances with the AOA and generates higher drag. This trend is the same in every case in our study. At 0.8% trailing edge thick condition, the pressure difference between the upper and lower edge is the maximum at the stall angle.

We must delve into the velocity distribution of the flow around the airfoil in order to adequately analyze the performance of the airfoil. Figures 8,9 and 10 demonstrate the velocity diagram of the 0% trailing edge thick, 0.2% trailing edge thick, and 0.8% trailing edge thick airfoil. In almost every case in our analysis, the upper edge of the airfoil has the higher velocity, whereas, the lower edge has a lower velocity as expected. However, as the value of AOA increases, the flow velocity steadily falls, and the velocity findings were relatively better in the lower section of the airfoil, which results in inferior performance overall restricting power generation. Flow separation is noticed at the higher angle of attack in almost every configuration in our analysis.

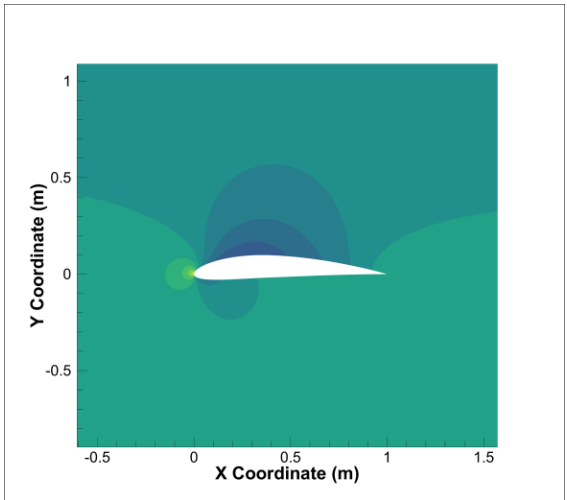
(a) α = 0 °

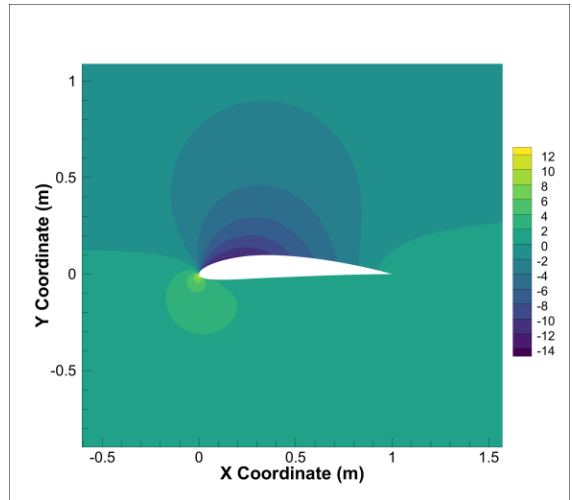
(b) α = 4 °

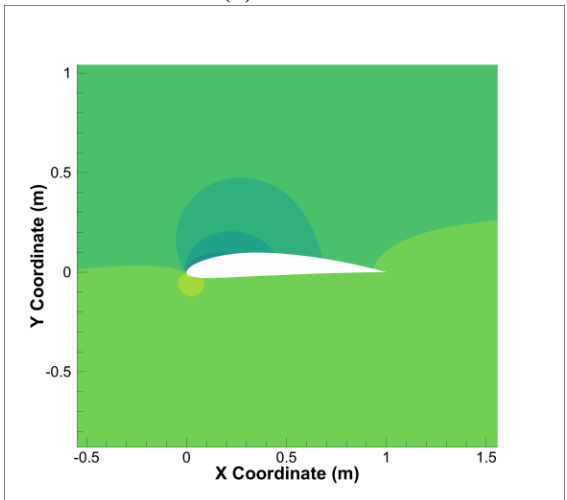
(c) α = 7 °

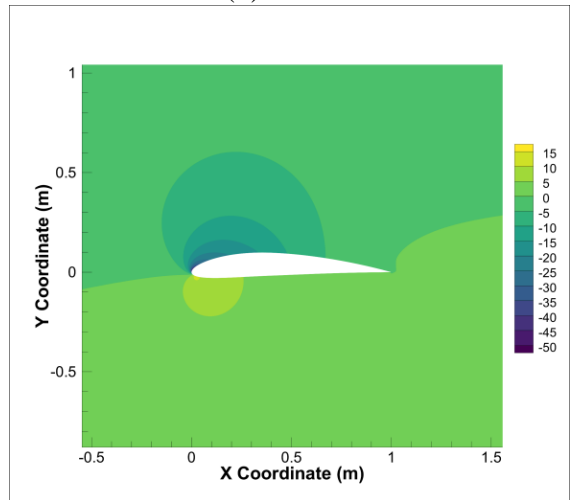
(d) α = 11 °

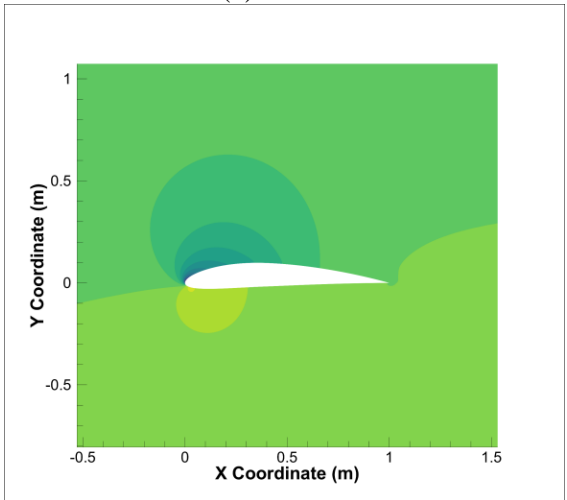
(e) α = 12 °

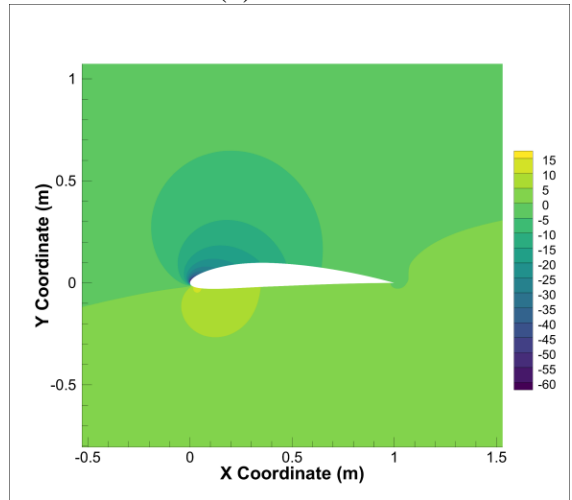
(f) α = 13 °

**Fig. 6.** NACA 4412 static pressure contours at 0 °, 4 °, 7 °, 11°, 12°, 13° angle of attack for 0% TE gap.

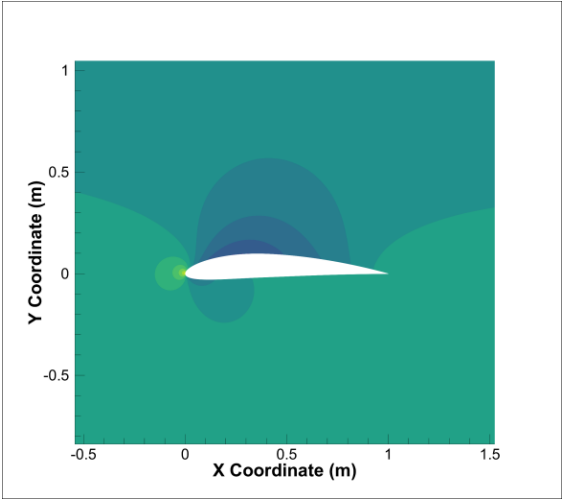
(a) α = 0 °

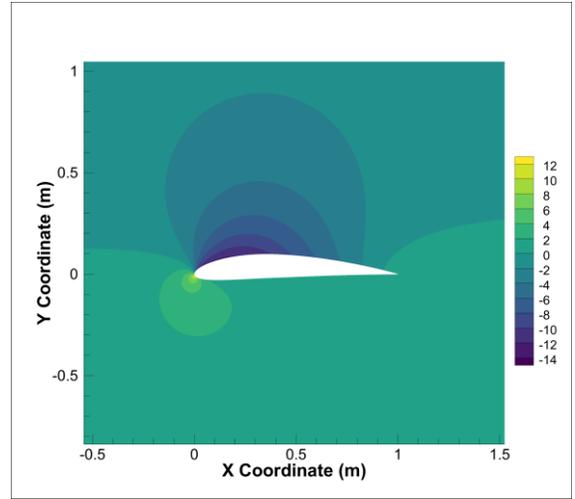
(b) α = 4 °

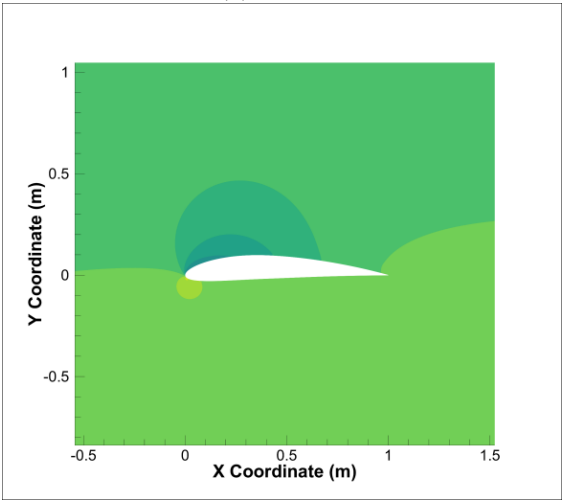
(c) α = 7 °

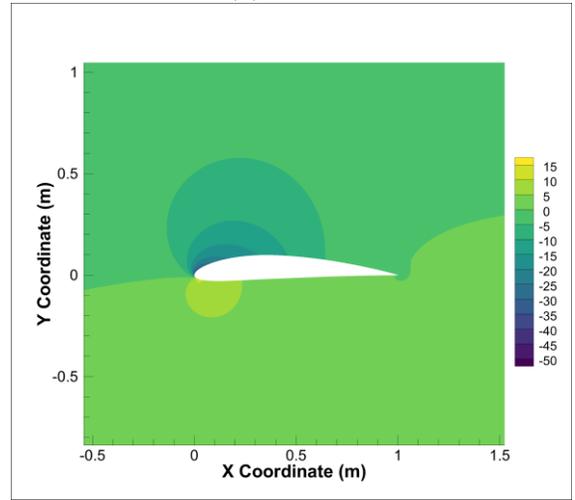
(d) α = 11 °

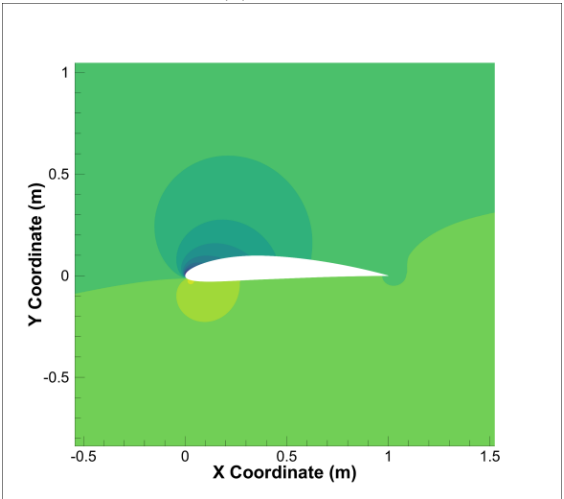
(e) α = 12 °

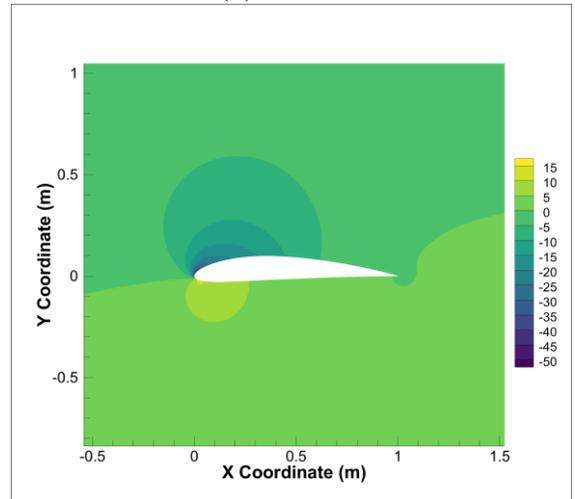
(f) α = 13 °

**Fig. 7.** NACA 4412 static pressure contours at 0 °, 4 °, 7 °, 11°, 12°, 13° angle of attack for 0.2% TET.

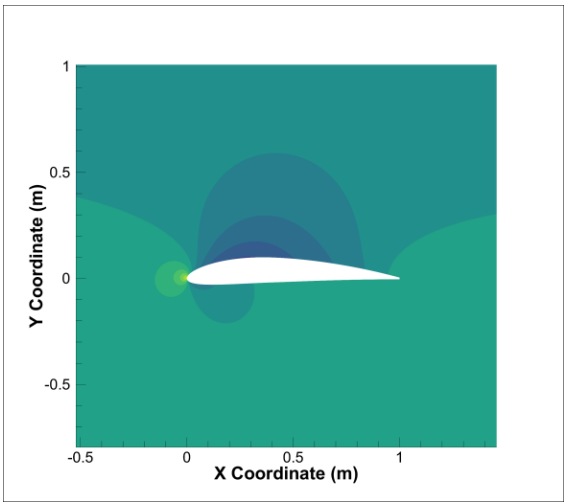
(a) α = 0 °

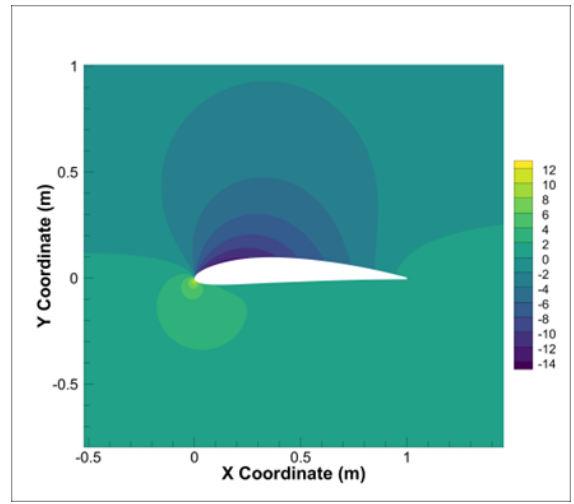
(b) α = 4 °

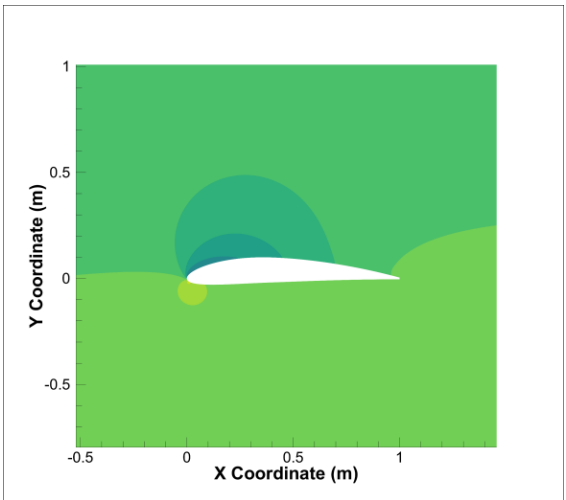
(c) α = 7 °

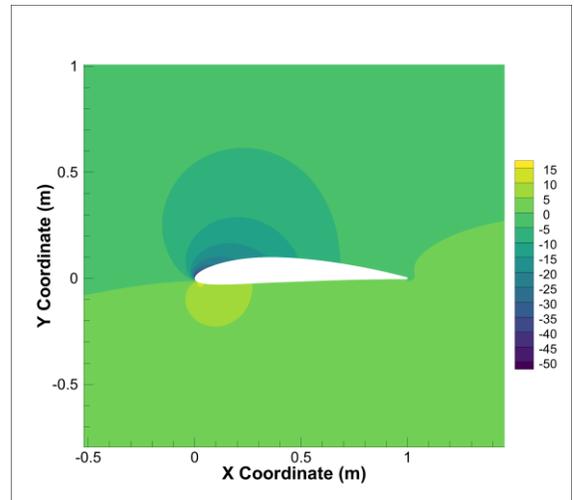
(d) α = 11 °

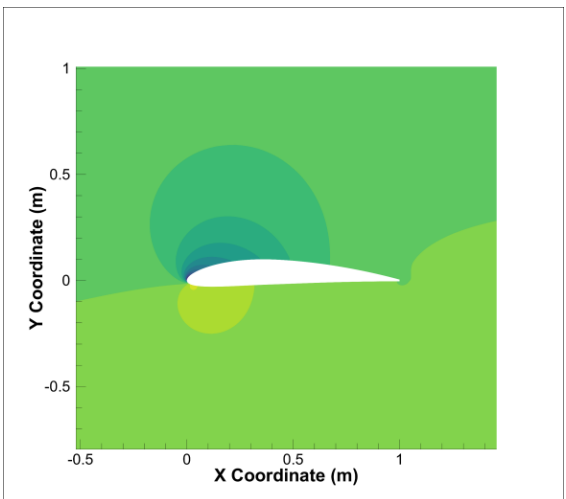
(e) α = 12 °

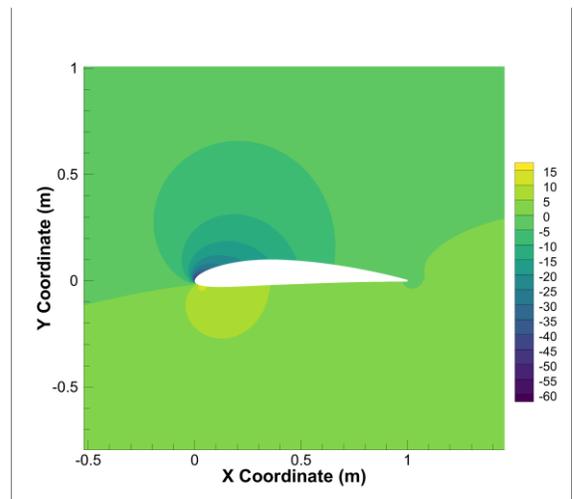
(f) α = 13 °

**Fig. 8.** NACA 4412 static pressure contours at 0 °, 4 °, 7 °, 11°, 12°, 13° angle of attack for 0.8% TET.

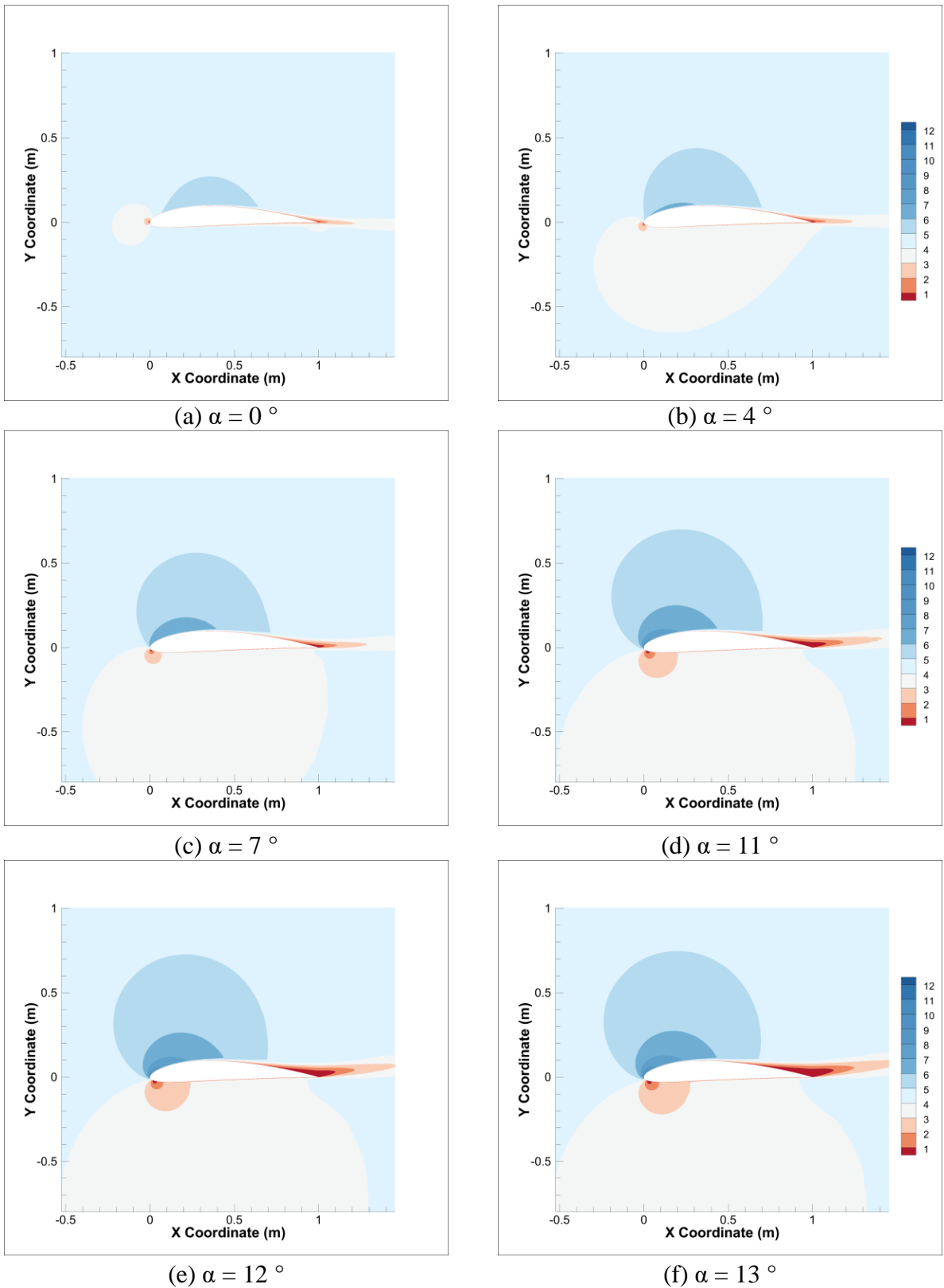

**Fig. 9.** NACA 4412 Velocity contours at 0 °, 4 °, 7 °, 11°, 12°, 13° angle of attack for 0% TET.

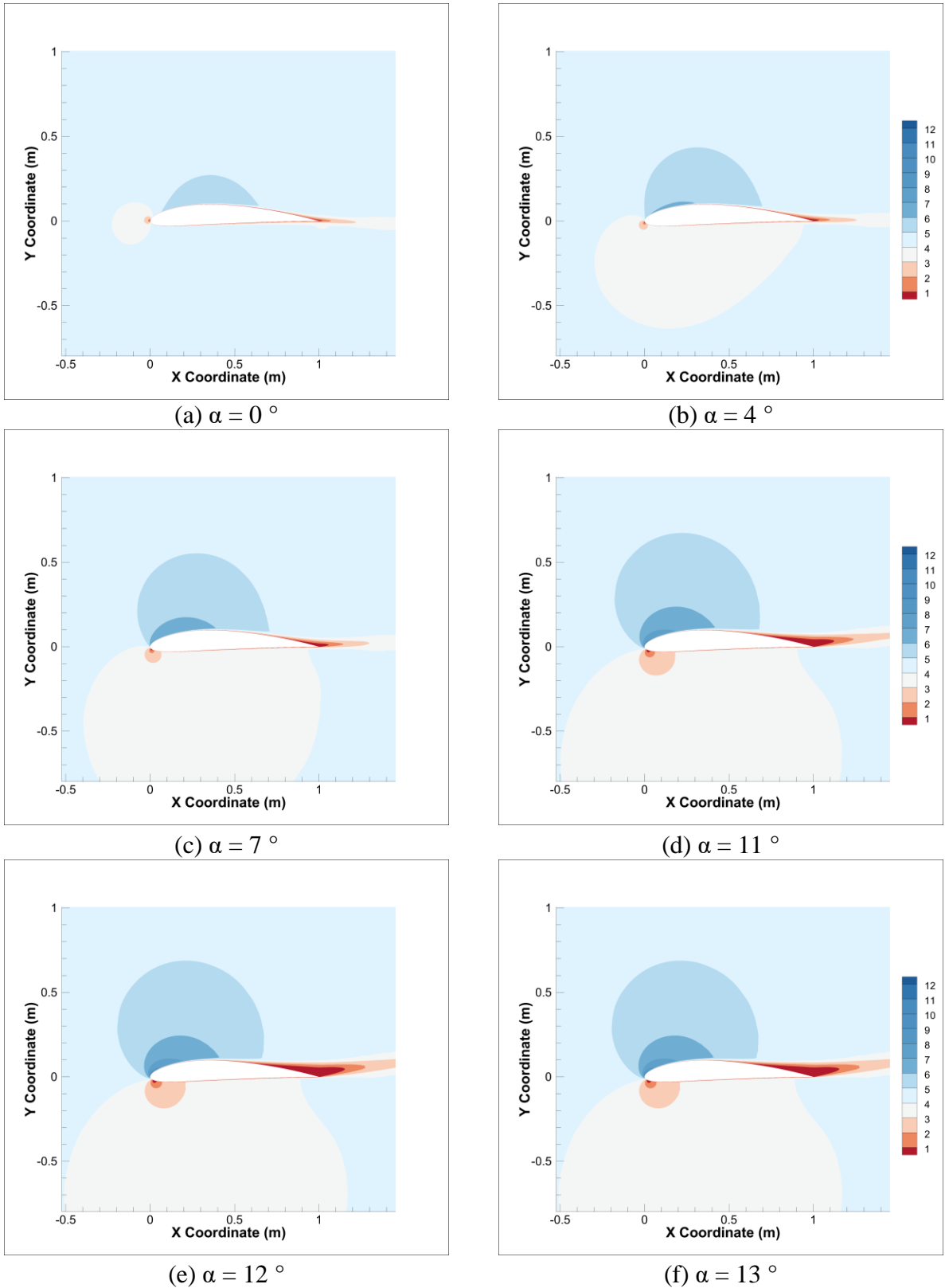

**Fig. 10.** NACA 4412 Velocity contours at 0 °, 4 °, 7 °, 11°, 12°, 13° angle of attack for 0.2% TET.

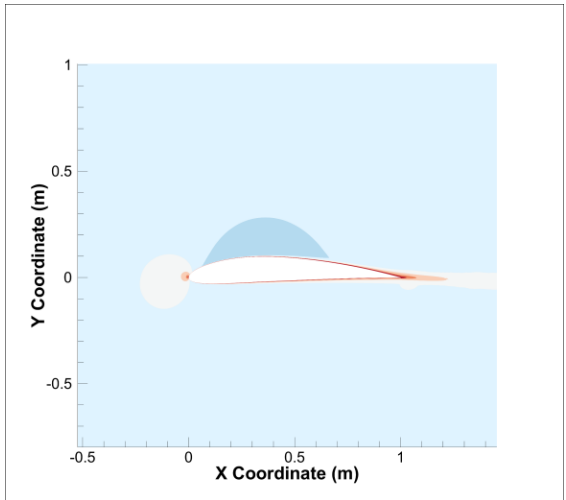
(a) α = 0 °

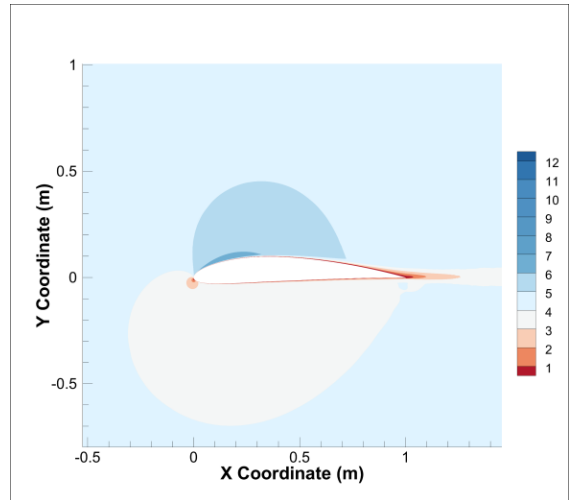
(b) α = 4 °

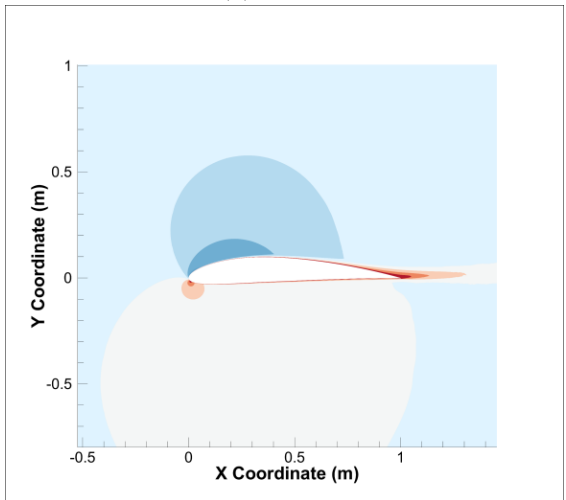
(c) α = 7 °

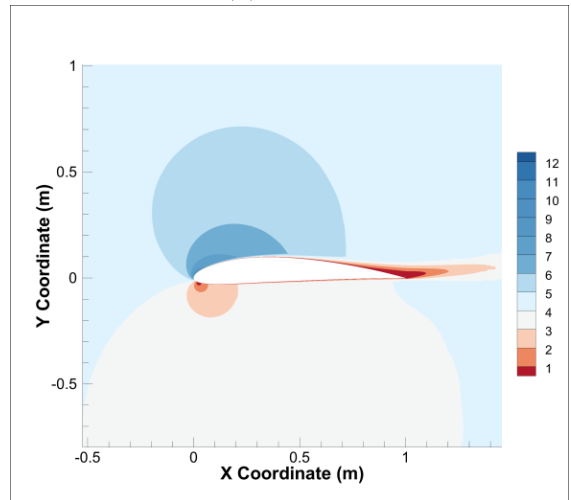
(d) α = 11 °

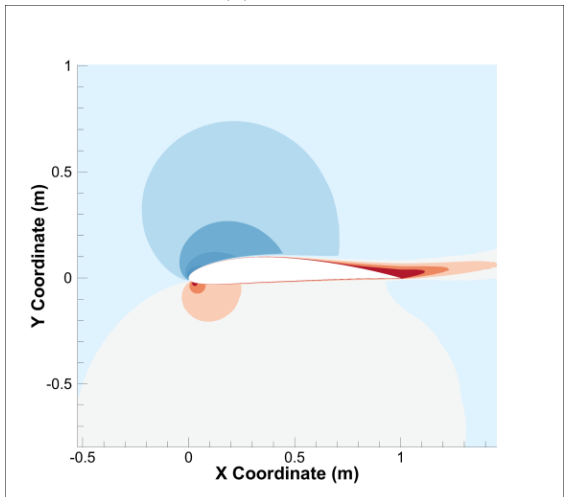
(e) α = 12 °

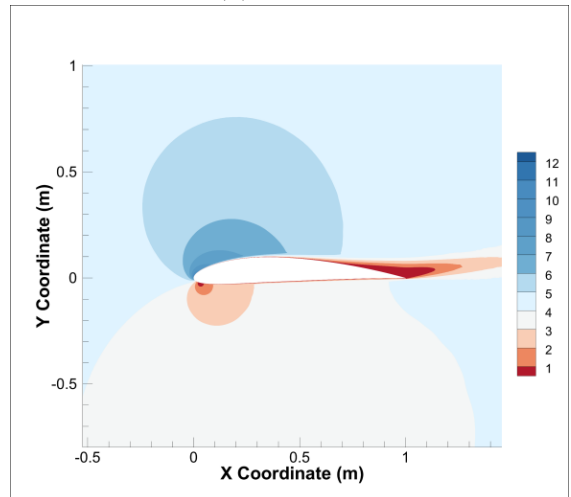
(f) α = 13 °

**Fig. 11**. NACA 4412 Velocity contours at 0 °, 4 °, 7 °, 11°, 12°, 13° angle of attack for 0.8% TET.

# 4. CONCLUSION

The major focus of this study is to understand the impact of the trailing edge thickness on the NACA 4412 airfoil. In this article, a variety of thicknesses on the trailing edge of NACA 4412 airfoil are taken into account, and we then attempt to identify the ideal trailing edge thickness that offers the highest performance among all thickness configurations. The Spalart-Allmaras turbulence model is selected for this study. The Reynolds number has been utilized in this study is 300k. To commence, 0% trailing edge thickness is taken for simulation. Then the trailing edge thickness has been started to enhance. For 0 to 1% thickness, the simulation is done with a 0.1% interval. Finally to recapitulate the research:

- The lift coefficient, drag coefficient, and lift coefficient-to-drag coefficient ratio are found at different angles of attack after simulation for these different trailing edge thickness conditions. Following the study, we have discovered that the optimum performance is achieved while considering the 0.8% thickness in the trailing edge because it displays a better $C_L/C_D$ vs AOA curve among all conditions. And, the 0.2% trailing edge thickness demonstrates the worst performance.
- According to the coefficient of lift coefficient scenario, almost every airfoil's lift coefficient increases with the AOA up to 12° or 14°, or the "stall angle," when the value of $C_L$ is said to peak. The value of the coefficient of lift starts to decline after passing the stall angle. In our study, the optimal $C_L$ vs angle of attack diagram is found to be for the 0.8% trailing edge thickness. On the contrary, The NACA 4412 airfoil's 0.2% trailing edge thickness illustrates the worst scenario of $C_L$ vs AOA.
- In the case of the drag coefficient vs AOA, it is clear that every airfoil, which is simulated in this study, experiences an increase in drag as the AOA increases. When the AOA is close to 20°, the value of the coefficient of drag is much higher in every case, in spite of the fact that it grows slowly with the angle of attack at the initial level.
- Even though the higher coefficient of lift is found at 12° to 14° angle of attack in almost every case in our study, the overall performance is lower because of the higher drag coefficient at this position. In general, we find the maximum $C_L/C_D$ vs AOA at 6° to 8° AOA in all 11 conditions.

- In general, the maximum $C_L$ is found at the higher AOA, however, the overall performance is not sufficient enough because of the higher $C_D$.
- On the static pressure analysis, it is found that the pressure difference of both sides of the airfoil enhances till stall angle and this difference is the highest on the stall angle. And pressure difference on the upper and leading edge of the airfoil increases with the angle of attack, which is responsible for the higher drag.